\documentclass[twocolumn,trackchanges]{aastex701}
\usepackage{xspace}
\usepackage{amsmath}
\usepackage{comment}

\newcommand{\ha}{H$\alpha$}
\newcommand{\lya}{Ly$\alpha$}
\newcommand{\pab}{Pa$\beta$}
\newcommand{\pag}{Pa$\gamma$}
\newcommand{\OII}{[\textrm{O}~\textsc{ii}]\xspace}
\newcommand{\OIII}{[\textrm{O}~\textsc{iii}]\xspace}
\newcommand{\Hb}{\textrm{H}\ensuremath{\beta}\xspace}
\newcommand{\oi}{[O\thinspace{\sc i}]}
\newcommand{\sii}{[S\thinspace{\sc ii}]}
\newcommand{\nii}{[N\thinspace{\sc ii}]}
\newcommand{\hei}{He\thinspace{\sc i}}
\begin{document}

\title{A Single Line-Emitter Identified in an Unbiased Search with JWST/NIRSpec:  Multi-Object Coronagraphy for Slitless Spectroscopy}

\author[orcid=0009-0002-9932-4461]{Mason S. Huberty}
\affiliation{Minnesota Institute for Astrophysics, University of Minnesota, 116 Church Street SE, Minneapolis, MN 55455, USA}
\email[show]{huber458@umn.edu}  

\author[0000-0002-9136-8876]{Claudia Scarlata}
\affiliation{Minnesota Institute for Astrophysics, University of Minnesota, 116 Church Street SE, Minneapolis, MN 55455, USA}
\email{mscarlat@umn.edu} 

\author[0000-0001-8587-218X]{Matthew J. Hayes}
\affiliation{Stockholm University, Department of Astronomy and Oskar Klein Centre for Cosmoparticle Physics, AlbaNova University Centre, SE-10691, Stockholm, Sweden}
\email{matthew@astro.su.se}   

\author[0000-0002-6586-4446]{Alaina Henry}
\affiliation{Space Telescope Science Institute, 3700 San Martin Drive, Baltimore, MD 21218, USA}
\email{ahenry@stsci.edu} 

\author[0000-0002-4153-053X]{Danielle A. Berg}
\affiliation{Department of Astronomy, University of Texas at Austin, 2515 Speedway, Austin, TX 78712, USA}
\email{daberg@austin.utexas.edu}

\author[0000-0002-2070-9047]{Vieri Cammelli}
\affiliation{Department of Physics, Informatics \& Mathematics, University of Modena \& Reggio Emilia, via G. Campi 213/A, 41125, Modena, Italy}
\email{vieri.cammelli@unimore.it}

\author[0000-0001-7782-7071]{Richard Ellis}
\affiliation{Department of Physics and Astronomy, University College London, Gower Street, London WC1E 6BT, UK}
\email{richard.ellis@ucl.ac.uk}

\author[0000-0002-6790-5125]{Anne E. Jaskot}
\affiliation{Department of Physics and Astronomy, Williams College, Williamstown, MA 01267, USA}
\email{08aej@williams.edu}

\author[0000-0001-8792-3091]{Yu-Heng Lin}
\affiliation{Caltech/IPAC, 1200 E. California Boulevard, Pasadena, CA 91125, USA}
\email{h} 

\author[0000-0001-7166-6035]{Vihang Mehta}
\affiliation{IPAC, California Institute of Technology, 1200 E.~California Blvd, Pasadena, CA 91125, USA}
\email{vmehta@ipac.caltech.edu}

\author[0000-0003-2589-762X]{Matilde Mingozzi}
\affiliation{AURA for ESA, Space Telescope Science Institute, 3700 San Martin Drive, Baltimore, MD 21218, USA}
\email{mmingozzi@stsci.edu}

\author[0000-0003-2083-7564]{Pierluigi Monaco}
\affiliation{Dipartimento di Fisica - Sezione di Astronomia, Università di Trieste, Via Tiepolo 11, 34131 Trieste, Italy}
\affiliation{NAF-Osservatorio Astronomico di Trieste, Via G. B. Tiepolo 11,
34143 Trieste, Italy}
\email{pierluigi.monaco@inaf.it}

\author[0000-0001-8419-3062]{Alberto Saldana-Lopez}
\affiliation{Stockholm University, Department of Astronomy and Oskar Klein Centre for Cosmoparticle Physics, AlbaNova University Centre, SE-10691, Stockholm, Sweden}
\email{alberto.saldana-lopez@astro.su.se}

\author[0000-0001-5333-9970]{Aayush Saxena}
\affiliation{Department of Physics, University of Oxford, Denys Wilkinson Building, Keble Road, Oxford OX1 3RH, UK}
\email{aayush.saxena3009@gmail.com} 

\author[orcid=0000-0002-3389-9142]{Jonathan C. Tan}
\affiliation{Dept. of Space, Earth \& Environment, Chalmers University of Technology, Chalmersgatan 4, 412 96 Gothenburg, Sweden} 
\affiliation{Department of Astronomy, University of Virginia, 530 McCormick Rd, Charlottesville, VA 22904, USA}
\email{jctan.astro@gmail.com}

\author[0000-0001-9136-3701]{Alice Young}
\affiliation{Stockholm University, Department of Astronomy and Oskar Klein Centre for Cosmoparticle Physics, AlbaNova University Centre, SE-10691, Stockholm, Sweden}
\email{alice.young489@gmail.com}

\begin{abstract}
We report the detection of a previously unknown object with a single emission line within the Hubble Ultra Deep Field, using a new JWST NIRSpec technique called Multi-Object Coronagraphy for Slitless Spectroscopy (MOCSS) with the G140M/F100LP observing mode. The MOCSS technique involves opening almost all the shutters on the NIRSpec microshutter array, while simultaneously closing the shutters that lie on top of known low redshift objects, facilitating a wide field slitless spectroscopic search without contamination from bright objects. 
This observational technique has enabled the detection of a strong emission line in four different dither positions and in two position angles, and thus cannot be artificial. This line is detected at 1.60 $\rm \mu m$ with a line flux of $[2.5\pm 0.3]\times10^{-18} \rm erg\ s^{-1} cm^{-2}$ and a FWHM of $13.2\pm2.0$\AA, and does not correspond to any known source with some of the deepest HST and JWST imaging in existence. 
Based upon the absence of imaging detection, we place a lower limit on the observed frame equivalent width (EW) of 7620\AA\ on this object, which strongly limits the number of possible lines that this line can be. 
Given the wavelength, an EW limit that is near the ceiling of expectation, and the lack of other lines in the bandpass, this line cannot be easily explained by \OII\ at $z=3.29$, \OIII\ at $z=2.20$, \ha\ at $z=1.44$, and \lya\ at $z=12.16$. 
We discuss possible AGN and transient phenomena that could explain the finding. 
\end{abstract}

\section{Introduction}
The James Webb Space Telescope (JWST) has ushered in a new era in the search and study of the furthest and faintest objects in the Universe.
Spectroscopically identified ultraviolet (UV) emission lines in these high-z sources can constrain both the conditions of the interstellar medium in the first galaxies and determine the contribution of faint AGN to the galaxy population \citep[e.g.,][]{castellano2022,curti2023,tang2023,fujimoto2023,llerena2025}. UV emission lines are important diagnostics of the star-formation, gas density, metallicity, dust content, the baryon cycle, and ionization properties of galaxies \citep[e.g.,][]{vidalgarcia2017,byler2018,kewley2019,llerena2022,mingozzi2024,hayes2025}.

Wide Field slitless spectroscopy (WFSS) observing modes have enabled the discovery of previously unknown emission line sources as far back as the Epoch of Reionization (EoR) \citep[e.g.,][]{treu2022,matthee2023,malkan2025,sun2025}. 
The Near-Infrared Imager and Slitless Spectrograph (NIRISS) WFSS observing mode does just this, but only has a resolution of $\rm R\sim100$, meaning it cannot capture emission lines with EW below $\sim10$\AA. The Near Infrared Camera (NIRCam) WFSS observing mode has a higher resolution of $\rm R\sim1000$, but at $\lambda>2.4$ $\rm \mu m$. 
Unfortunately, it is the $1<\lambda<2.5$ $\rm \mu m$ regime that hosts $z>6$ UV emission lines. 
In this wavelength range, the highest resolution ($R\gtrsim1000$ with G140M/H) spectroscopic instrument is the Near Infrared Spectrograph \citep[NIRSpec, ][]{jakobsen2022}.
However, typical NIRSpec observations require a pre-selection of targets unlike slitless spectroscopic modes, which introduces biases to the spectroscopic samples based upon magnitudes and/or colors, and prohibits the identification of unknown emission line objects. Therefore, we have pioneered a new NIRSpec technique, Multi-Object Coronagraphy for Slitless Spectroscopy (MOCSS), where we open all shutters in the microshutter assembly (MSA) except those on known low-z targets, thereby obtaining WFSS spectroscopy without contamination from unwanted bright objects.
This allows us to search for never before seen faint and high-$z$ emission lines without object pre-selection, at $\rm R\sim1000$ in the $\sim1-2$ $\rm \mu m$ wavelength range. The lack of pre-selection bias is a very important characteristic for such a survey to have, so that the most recent puzzling discoveries of the high-z Universe are not subject to some kind of observational bias.

In this paper, we present a previously unknown line-emitter within the MOCSS Hubble Ultra Deep Field (HUDF) survey, corresponding to no known source. 
In section~\ref{sec:observations}, we summarize the MOCSS HUDF observation strategy and data reduction process. In section~\ref{sec:methodology}, we present the methodology for identifying the line-emitter. In section~\ref{sec:results}, we present the emission line and discuss the possible origins and implications of the line. Our conclusions are outlined in section~\ref{sec:conclusion}. 
We assume a flat $\Lambda$CDM Planck cosmology \citep{planck2020} and use the AB magnitude system \citep{oke1983}.

\section{Observations and Data Reduction}\label{sec:observations}
The Hubble Ultra Deep Field \citep[HUDF, overlapping with the Great Observatories Origins Deep Survey-South, GOODS-S,][]{beckwith2006,whitaker2019} is a well-studied field, known for its exceptional wavelength coverage and depth of observations. 
This includes 11-band Hubble Space Telescope (HST) imaging \citep[][]{beckwith2006,bouwens2010,teplitz2013,ellis2013,hayes2024} and both shortwave and longwave imaging from the JWST NIRCam \citep{williams2023,eisenstein2023} and NIRISS \citep[NGDEEP,][]{bagley2024}. 

The Multi-Object Coronagraphy for Slitless Spectroscopy (MOCSS) HUDF survey, JWST-3290 (PIs: Hayes \& Scarlata), was conducted in JWST Cycle 2. 
The primary overview of the MOCSS survey is discussed in \citet{huberty2026mocssoverview}. Here we present a summary of the MOCSS observation strategy and the data reduction process.

MOCSS observations were obtained with NIRSpec G140M/F100LP disperser/filter pairing, which nominally covers the wavelength range between 0.97 and $\sim1.89$ $\rm \mu m$ at $\rm R\sim1000$. Instead of using the standard Multi-Object Spectroscopy (MOS) observing mode, all shutters were left open to enable a slitless spectroscopic campaign.

To minimize contamination from bright, low-z sources
and because the primary goal of the MOCSS survey is to detect faint emission lines in the EoR ($z\gtrsim6.9$), we masked all sources with with flux detected at $\lambda<9600$\AA, identified in HUDF HST F814W images \citep{whitaker2019}, by closing shutters that lie on top of these objects.
Accounting for failed shutters and intentionally closed shutters to mask $z<6.9$ galaxies, $\gtrsim 90 \%$ of the MSA area was left open.

The HUDF was observed at two orientations (offset by $\sim89^\circ$), with 8 separate exposures per orientation (for a total of 16 exposures). 
We created a dither pattern optimized to ensure uniform illumination of the field. The dither pattern consists of 4-point dither pattern of half a shutter, followed by one large shift ($\sim21$") to cover the NIRSpec chip gap where a second 4-point dither pattern of half a shutter each takes place.

\begin{table*}[]\label{tab:dithers}
\centering
\caption{Details of the MOCSS observations. Information about each exposure in the dither pattern. Each group of four exposures are aligned and stacked, resulting in stacks A, B, C, and D. The right ascension/declination offsets are calculated with respect to exposure 1 for all orientation 1 objects, and exposure 9 for all orientation 2 objects.}\label{tab:dithers}
\begin{tabular}{cccccc}
\hline
Exposure             & Orientation          & Stack & Right Ascension Offset & Declination Offset     \\
\multicolumn{1}{l}{} & \multicolumn{1}{l}{} &\multicolumn{1}{l}{} & ($^{\prime\prime}$)         & ($^{\prime\prime}$)               & \multicolumn{1}{l}{} \\ \hline
1                    & 1         & A           & 0                  & 0                                        \\
2                    & 1          & A         & 0.26                             & 0.81                                                 \\
3                    & 1          & A          & 0.52                   & 1.61                                        \\
4                    & 1         & A           & 0.77                   & 2.42                                       \\
5                    & 1         & B          & -15.97                  & 14.62                                      \\
6                    & 1          & B          & -15.71                         & 15.42                                               \\
7                    & 1          & B          & -15.45                  & 16.23                                      \\
8                    & 1         & B           & -15.20                 & 17.03                                  \\
\hline
9                    & 2        & C            & 0                 & 0                                 \\
10                   & 2         & C           & 0.91                             & -0.24                                           \\
11                   & 2          & C          & 1.81                  & -0.48                                     \\
12                   & 2         & C           & 2.72                   & -0.72                                  \\
13                   & 2          & D          & 16.79                  & 13.88                             \\
14                   & 2           & D         & 17.70                         & 13.64                                          \\
15                   & 2          & D          & 18.60                  & 13.40                                       \\
16                   & 2          & D          & 19.51                 & 13.15                                \\ \hline
\end{tabular}
\end{table*}

Each of the 16 exposures was row-by-row sky subtracted using a custom algorithm. This is required because in each row, the value of the background depends on the number of shutters opened in that specific row (which varies depending on the number of shutters that were closed to mask sources in each row). 
The 16 exposures were divided into 4 groups (called groups A, B, C, and D) corresponding to each of the small dither groups/orientations (see Table~\ref{tab:dithers} for dither information). 
Dispersed images were aligned relative to the first exposure within each group (e.g., exposure 5 for group B) to ensure proper alignment of spectral traces across all exposures.
We then created deep dispersed images for each group by error-weighted-mean-combining these aligned images.
In this process, for each group we also obtain the standard deviation per aligned pixel into a $\sigma$ map.

The results of this procedure are a set of 4 dispersed, sky subtracted images that we use in the next section to detect emission lines.

\begin{figure*}
    \centering
    \includegraphics[width=.9\linewidth]{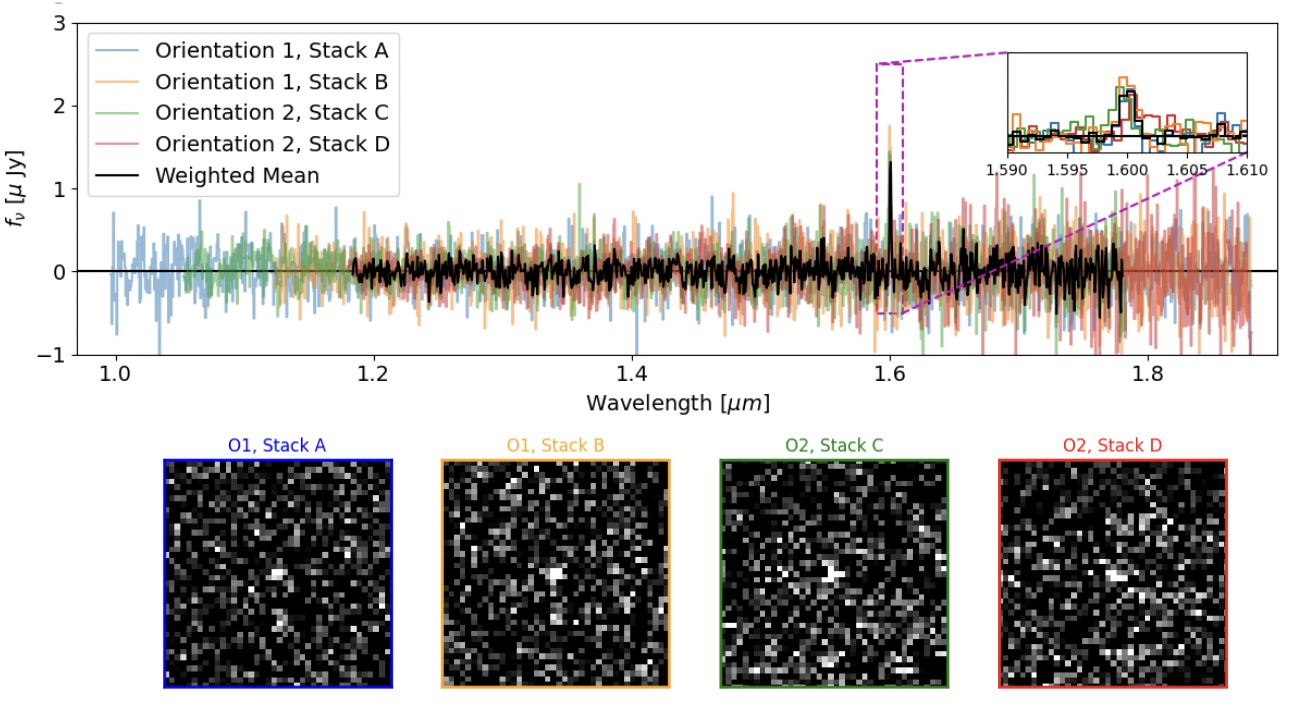}
    \caption{Single emission line candidate matched in all stacks and orientations. Top: The 1D spectrum of the line in the four stacks. The black line is the weighted-mean of the four individual stacks (and is only plotted in the wavelength range where all four stacks overlap). The inset zooms in on the location of the emission line. Bottom: The direct, \textsc{sep}-identified detector images of the line in each of the four stacks (where O1=orientation 1, O2=orientation 2).
    }
    \label{fig:mocssspectralextraction}
\end{figure*}
\section{Methodology}\label{sec:methodology}
In standard slitless spectroscopy, direct images of the field are used to extract the spectra of known sources. In this work, we instead identify emission lines directly in the dispersed images \citep[as done, e.g., in][]{runnholm2025,kashino2023}.

\subsection{Emission Line Detection}\label{sec:sep}

Candidate emission lines in the dispersed images were identified using the Python library for Source Extraction and Photometry (\textsc{sep}) package \citep{barbary2016,bertin1996}. 
We use \textsc{sep.Background} to create a spatially variable background image using $32\times32$ pixel background boxes. 
We retain only sources with more than 5 contiguous pixels with flux above a threshold of $3\sigma_{rms}$. Each detector stack includes $\sim200-300$ candidate emission lines.

Real emission lines are expected to move in a very specific way between different stacks. Within a single orientation, real lines would appear in both stacks, offset by the large dither that corresponds to about 200 pixels in the dispersion direction and 2 in the cross-dispersion direction. This shift may not be exact across the entire field of view due to  wavelength- and position-dependent distortions that are not fully captured by the dither correction and can introduce additional offsets. We visually inspect the identified candidates, to ensure that they are not located near the edge of the detector and exhibit the morphology expected for emission lines.
We classify sources as genuine emission line detections
if their centroids in the two groups lie within 4 pixels of the expected position.
This procedure yields 21 and 21 emission-line candidates for orientations 1 and 2, respectively (for a total of 42).

\subsection{Matching Emission Line Candidates in Different Orientations}\label{sec:matching}

The emission lines identified in the two orientations and described in the previous section do not have unique wavelength assignments. This ambiguity is the reason the field was observed at two different dispersion angles. In slitless observations, the wavelength corresponding to a given pixel in the dispersed image depends on the position of the source within the field of view. Consequently, the same pixel can be associated with different wavelengths for different objects. 
A selected emission line target should correspond to a single source position and wavelength solution that is consistent across both orientations. We applied this criterion to the lists of emission line candidates identified in Section~\ref{sec:sep}.
Specifically:
\begin{enumerate}
    \item For each orientation, we mapped each emission line candidate to every possible MSA shutter that could host the source of that line. Note that each host shutter corresponds to a different wavelength for the line.
    \item We retain in the final list of secure lines any lines that map to the same shutter in both orientations, with wavelength agreement within 6\AA\ (i.e., roughly the G140M's dispersion)

\end{enumerate}

\begin{figure}
    \centering
    \includegraphics[width=.9\linewidth]{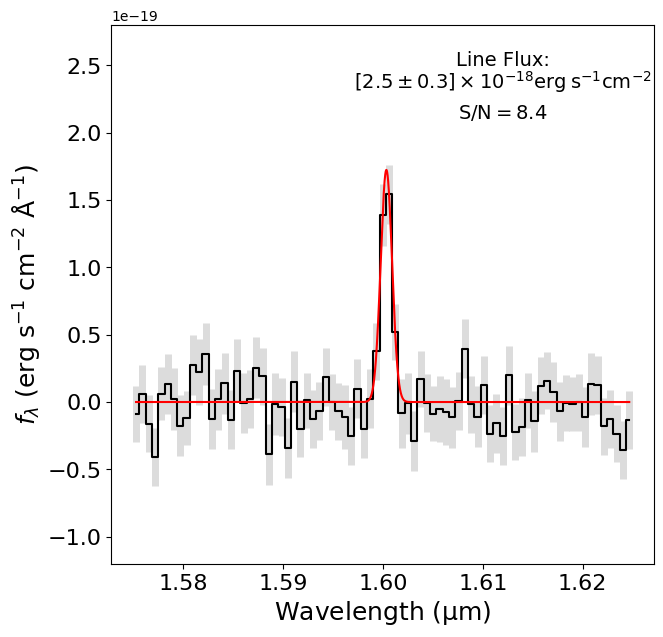}
    \caption{Weighted-mean-line of the four stacks (black) with a Gaussian fit to the emission line (red). The associated uncertainty on each point is highlighted in gray. This strong emission line peaks at 1.60 $\rm \mu m$, and corresponds to the location of R.A.=$53.13127^\circ$ and Decl.$=-27.78908^\circ$.
    }
    \label{fig:medianline}
\end{figure}

\subsection{Spectral Extraction}\label{sec:spectralextraction}
The criteria outlined in the previous section yield seven secure emission-line candidates. 
Six of the seven emission lines have well-defined hosts; galaxies identified in the JWST Advanced Deep Extragalactic Survey \citep[JADES,][see section~\ref{sec:lackofhost}]{eisenstein2023}. In addition to these sources, many of the remaining emission lines not seen in both orientations are constrained from being multi-line emitters, from being found in ancillary spectroscopy, and from being detected in 3 out of 4 stacks (with the emission line in the remaining stack falling in the chip gap, out of wavelength coverage, or falling just below the \textsc{sep} detection threshold). In total $\gtrsim75\%$ of the 42 emission lines are robustly identified and have known hosts.
These sources are discussed in \citet{huberty2026mocssoverview}.

However, there is one remaining emission line that is seen in both orientations that has no identifiable host in JADES and originates from a sky position of R.A.=53.13127$^\circ$ and
Decl.=-27.78908$^\circ$. 
This position is confirmed post-spectral extraction from the pixel to R.A./Decl./wavelength transformation in the extracted 2D spectra.  
This position corresponds to an emission-line wavelength of $16004\pm1 \rm \AA$.

To search for secondary emission lines associated with this object, we extracted the spectra at both orientations by running the JWST stage~2 pipeline on the stacked dispersed images. 
The pipeline version used in this extraction was the JWST v1.19 pipeline \citep{jwstpipeline2023}. We used the jwst\textunderscore 1413.pmap reference file in the JWST Calibration References Data System \citep[CRDS,][]{jwstcrds2016}  and included the latest S-flat images (released summer 2025).
The stage 2 pipeline applies instrumental corrections and calibrations to the count-rate products that result in fully calibrated exposures. The pipeline steps of \textsc{master\textunderscore background, pathloss,} and \textsc{barshadow} are skipped, as they are either not required or would result in inaccurate corrections given our observing strategy.

The JWST spectroscopic flux calibration is computed from observations of a point source and includes an aperture correction for flux loss due to the finite shutter size in the dispersion direction. Our observations, however, do not suffer from this limitation (as neighboring shutters are opened along the dispersion direction). Accordingly, the flux calibration should be corrected to account for only the fraction of light lost because of the shutter bars. 
To estimate this number, we used the Python package STPSF \citep{perrin2014} to compute a model PSFs at 1.6$\mu$m and then simulated the spectrum of a point source through 3 adjacent shutters. We find that the flux loss is only 3.3\%.

The extracted 1D spectra in each of the 4 stacks can be seen in Figure~\ref{fig:mocssspectralextraction}.
The top panel shows the full 1D spectral extraction while the inset zooms in on the location of the emission line at 1.60 $\rm \mu m$.  
We also compute, in each wavelength bin, the weighted mean of the spectra from the four stacks. The resulting spectrum is shown as the black curve in Figure~\ref{fig:medianline}.
We generate 1000 realizations of the spectra from each of the four stacks by drawing values from a Gaussian  distribution centered at $f_\nu$ and a standard deviation equal to the error on $f_\nu$. We take the median of the four spectra (one per each stack) in every wavelength bin for all 1000 realizations; the standard deviation of the median distribution in each wavelength bin is taken as the error on the final spectrum.
The bottom row of Figure~\ref{fig:mocssspectralextraction} shows cutouts of the dispersed images centered at the emission line position in each stack. 

\subsection{Emission Line Measurement}
We fit a Gaussian to the detected emission line with \textsc{scipy.optimize.leastsq} \citep{Scipy:2020} using a Levenberg-Marquardt algorithm.
A Monte Carlo (MC) approach is used to estimate the error on the line flux: 
first, we created 1000 realization of the spectrum by drawing values from a Gaussian  distribution centered at $f_\lambda$ and a standard deviation equal to the error on $f_\lambda$. Then, we refit the Gaussian model to the 1000 realization of the spectrum and computed the error on the line flux as the dispersion of the derived line flux distribution. 
This measurement yield an integrated line-flux of $[2.5\pm 0.3]\times10^{-18} ~ \rm erg \: s^{-1}cm^{-2}$. We also measure the line flux in each stack individually, which is presented in Table~\ref{tab:indivudalfluxes}.
The same MC Gaussian fitting procedure is used to derived the Full Width Half-Maximum (FWHM) of the line, found to be $13.2\pm2.0$\AA\ in the observed frame.

We can use the line flux measurement together with the noise level to estimate a lower limit on the line Equivalent Width (EW). Measuring the noise as the standard deviation of the weighted-mean spectrum in the wavelength regime covered by all four stacks, we find that $\sigma_{f_\lambda}=2.0\times10^{-20} \rm erg \: s^{-1}cm^{-2}$\AA$^{-1}$. This corresponds to a 3$\sigma$ lower limit on the observed line EW$>$41\AA. As we will see in the next section, however, the non-detection in imaging could imply a much larger EW limit.

\begin{table*}[]
\centering
\caption{Properties of the emission line when measured individually in each stack and the combined stack. There is a time period of about 88 days between observations in orientation 1 (stacks A and B) and orientation 2 (stacks C and D). Given the lack of detection for this source in JADES deep NIRCam imaging, the $5\sigma$ upper limit on the line flux based on the non-detection would be $\sim1.1\times10^{-18} \rm erg\ s^{-1} cm^{-2}$.}\label{tab:indivudalfluxes}
\begin{tabular}{cccccc}
\hline
Stack & Flux & Significance     & FWHM & Central Wavelength & Date of\\
         & ($\rm erg\ s^{-1}cm^{-2}$)    &         &   (\AA)    &   (\AA)  & Observation \\ \hline
A            & $[2.3\pm0.5]\times 10^{-18}$           & $4.7\sigma$  & $10.7\pm2.9$ & $16006\pm2$ & 23 Aug 2023 \\
B            & $[3.5\pm0.6]\times 10^{-18}$           & $6.1\sigma$  & $14.5\pm3.2$ &  $16003\pm1$ & 24 Aug 2023 \\
C            & $[2.9\pm0.7]\times 10^{-18}$           &  $4.5\sigma$ & $14.9\pm5.4$ &  $16001\pm2$ &  19 Nov 2023 \\
D            &    $[2.9\pm1.2]\times 10^{-18}$        & $2.5\sigma$  & $23.3\pm14.4$ &   $16006\pm2$ & 19 Nov 2023 \\
\hline
Combined &    $[2.5\pm0.3]\times 10^{-18}$        & $8.4\sigma$  & $13.2\pm2.0$ &   $16004\pm1$ & \\

\hline
\end{tabular}
\end{table*}

\section{Discussion}\label{sec:results}
In this section we address the possible origin of the emission line.

\subsection{Emission Line Host, or Lack Thereof}\label{sec:lackofhost}
\begin{figure}
    \centering
    \includegraphics[width=.9\linewidth]{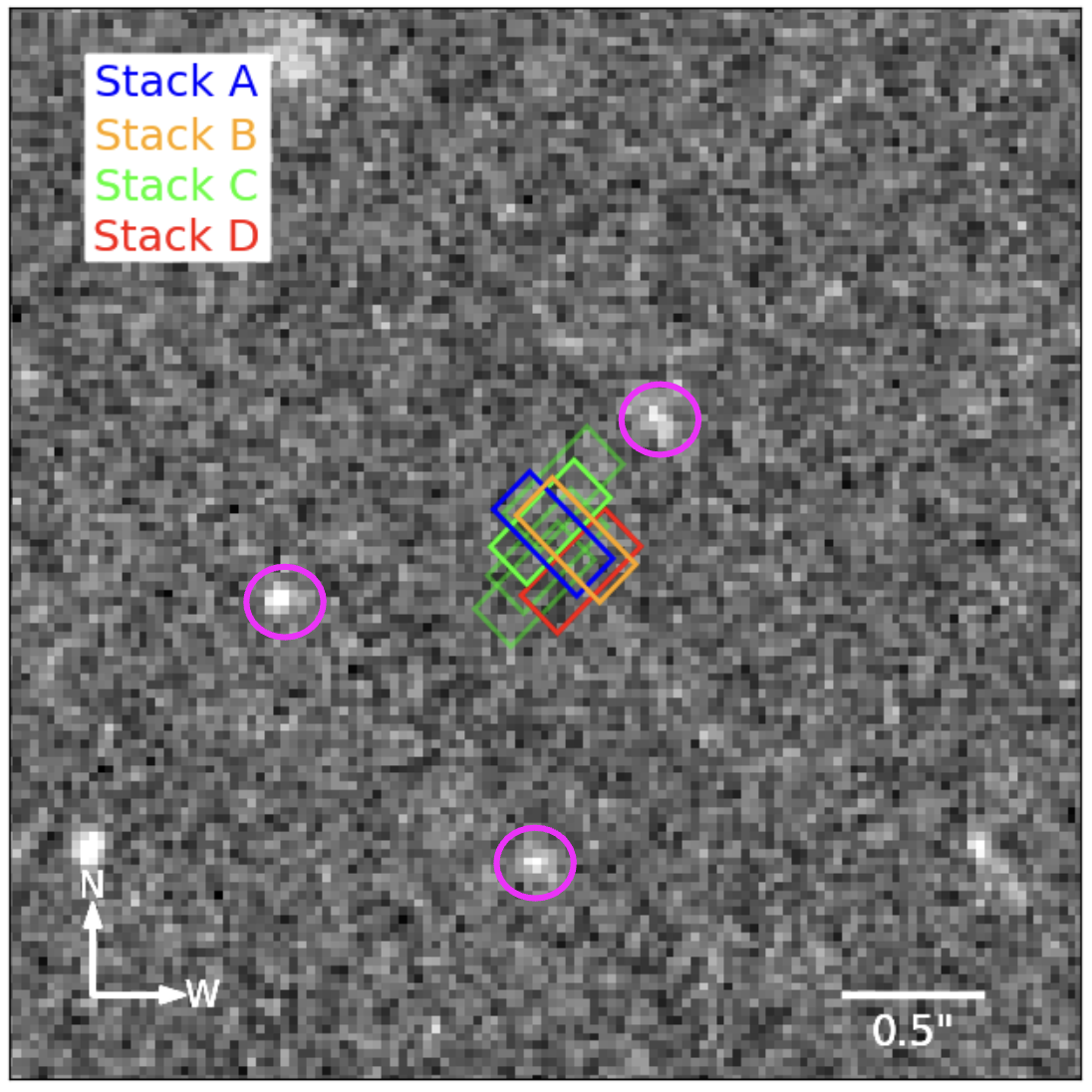}
    \caption{Location of the line-emitting source in NIRCam F150W imaging, with north at the top of the image, and east on the left side of the image. The source location corresponds to no known source in the JADES catalog. The pink circles indicate the nearest sources identified in JADES. The shutter imprints show the location of the reference shutter in each stack used for spectral extraction (blue for stack A, yellow for stack B, green for stack C, and red for stack D). To clarify that each shutter represents a stack, for stack C (green), we also plot faint shutters representing the three other exposures for that stack.
    }
    \label{fig:mysteryline}
\end{figure}

Figure~\ref{fig:mysteryline} shows a cutout of the JADES F150W NIRCam image centered at the inferred position of the source responsible for the emission line. The image reaches a $5\sigma$ depth of $4.6\rm nJy$ \citep[within a $0.3"$ diameter apertures, corresponding to a limiting $3\sigma$ magnitude of $30.3$,][]{rieke2023}. The figure also shows the positions of the reference shutters used for the spectral extraction in each of the four stacks. 
Although the green and red shutter appear more offset from one other than the blue and yellow shutters, this is a consequence of the source position inferred by the orientation matching code (Section~\ref{sec:matching}), which places the source near the boundary between two shutters. This does not affect our analysis because the dithering strategy causes the spectrum from each stack to sample flux from a larger region than that covered by the reference shutter alone. This is illustrated in Figure~\ref{fig:mysteryline}, where the faint green shutters indicate the positions of the other three dithers that contribute flux to the extracted spectrum for stack C.

It is clear that there is no source at the position of the shutters in the image shown in Figure~\ref{fig:mysteryline}, despite the image being among the deepest images ever obtained by JWST. Three sources (marked in pink in the figure) appear to be within one arcsecond from the putative position of the line host, but too far to be directly responsible for the emission line. In Sections~\ref{sec:agn} and~\ref{sec:transient} we discuss possible scenarios where these objects could still help the understanding of this line. 
Information about these neighboring sources can be found in Table~\ref{tab:nearbysources}. 

The lack of detection in the broad band image is surprising, because the F150W filter fully encompasses the wavelength of the emission line. Indeed the line flux alone would be sufficient to produce a detectable signal in the F150W observations. 
This non-detection places a strong lower limit on the  EW of the line.
The 5$\sigma$ depth of 4.6nJy corresponds to a $3\sigma$ continuum limit of $f_{\lambda}=3.23\times10^{-22} \rm erg\ s^{-1}cm^{-2}$\AA$^{-1}$. Combining this limit with the measured line flux yields a lower limit on the observed-frame equivalent width of 7620\AA.

\begin{table*}[]
\centering

\caption{The nearest JADES-identified sources to the line-emitter. These three source correspond to the pink circles in Figure~\ref{fig:mysteryline}. In Figure~\ref{fig:mysteryline}, the northwest JADES galaxy is to upper-right, the east galaxy is to the left, and the south galaxy is to the bottom of the figure. These neighboring galaxies could host an ionizing source that is illuminating a cloud of gas that is the source of the emission line. Sky coordinates are in epoch J2000.}\label{tab:nearbysources}
\begin{tabular}{ccccc}
\hline
Right & Declination & Location Relative     & z$_{phot}$ & F150W\\
   Ascension ($^\circ$)         & ($^\circ$)             &   to line-emitter    &  &  (nJy)   \\ \hline
$53.131165$            & $-27.788930$            & Northwest  & $0.22^{+3.11}_{-0.09}$ & $4.75\pm1.33$   \\
$53.131584$            & $-27.789105$            & East  & $6.03^{+0.04}_{-0.29}$ & $4.69\pm1.40$   \\
$53.131301$            & $-27.789367$            & South  & $0.19^{+2.60}_{-0.03}$ & $3.57\pm0.95$   \\
\hline
\end{tabular}
\end{table*}

\subsection{Interpretation as Star-Forming Galaxy}\label{sec:extragalactic}
The emission line at $1.60\ \rm \mu m$ cannot
be readily explained by any galactic source. 
If the emission line originates in an extragalactic source, its redshift depends on the line identification. Here we use constraints based on the absence of additional lines as well the inferred line EW and FWHM to explore possible strong line identifications. 

Figure~\ref{fig:mocssspectralextraction} shows that there are no additional emission lines in the wavelength range covered by the spectrum, which has a $3\sigma$ line flux limit that varies from $\sim8.2$ to $\sim10.6\times10^{-19} ~ \rm erg \: s^{-1}cm^{-2}$ depending on the wavelength. The constraints from the EW limit will depend on the redshift of the objects. 
In Table~\ref{tab:emissionlines}, we list strong emission lines that are commonly detected in the spectra of galaxies: \ha, \OIII, \OII, and \lya\ (no other potential emission lines theoretically get within a factor of 10 of the EW of this object). For each line, we compute and list the galaxy's properties that would result from that line identification. 
In the next sections, we discuss each possible identification individually.

\begin{table*}[]
\centering
\caption{Common strong emission lines and their respective rest wavelengths that could be the emission line found in the MOCSS survey. For each common emission line, we also list the corresponding redshift of the emission line source and the lower limit on the rest EW of the line (based on the lack of detection in imaging), the line luminosity, and rest FWHM should it match that emission line. The EW limits are very large given the lack of imaging detection baseline, near or above the theoretical and observational maximum in all cases, making it challenging for these emission lines to explain this emission line.}\label{tab:emissionlines}
\begin{tabular}{cccccc}
\hline
Emission Line & Rest Vacuum & $z$     & Rest EW Lower & Line Luminosity & Rest FWHM\\
   Candidates           & Wavelength (\AA)             &       & Limit (\AA)  &  $\log_{10}(\rm erg\ s^{-1})$  & (\AA)\\ \hline
\ha            & 6564.61            & 1.44  & 3130 & $40.5^{+0.1}_{-0.1}$  & $5.4\pm0.8$   \\
\OIII          & 5008.24            & 2.20  & 2380  & $41.0^{+0.1}_{-0.1}$  & $4.1\pm0.6$  \\
\OII           & 3729.88            & 3.29  & 1770 & $41.4^{+0.1}_{-0.1}$  & $3.1\pm0.4$ \\
\lya           & 1215.67            & 12.16 & 580 & $42.7^{+0.1}_{-0.2}$  & $1.0\pm0.1$   \\
\hline
\end{tabular}
\end{table*}

\subsubsection{Interpretation as \OII}
Should the line detected in the JWST spectrum be interpreted as the \OII\ doublet $3727,3729$\,\AA\ doublet\AA, the redshift would be $z_{[OII]}=3.29$ and $\rm EW_{OII}\gtrsim 1770$\AA. 
In the rest-frame, the \OII\ $3727,3729$\,\AA\AA, doublet is separated by $\sim2.7$\AA, or 11.6\AA\ in the observed frame, potentially resolvable in the $\approx 10$\AA\ resolution spectra. 
The doublet ratio, $R_{\mathrm{[O\,II]}} = \frac{I_{3729}}{I_{3726}}$, depends sensitively on the electron density, varying from $\sim 1.5$ in the low-density regime to $\sim 0.35$ in the high-density regime \citep{ferland2017,gunasekera2025}. Reproducing the narrow line width, therefore, requires high-density gas with $n_e \gtrsim 10^4\,\mathrm{cm}^{-3}$. At this density, [O\,II] will be suppressed by collisional de-excitation of $\rm O^+$ ions. 
In addition, the implied rest-frame equivalent width is extremely large, exceeding a few hundred~\AA, which has not been observed in [O\,II] emitters \citep[e.g.,][]{hogg1998,yan2006,ly2012,cedres2021}. Thus, the properties of this emission line cannot easily be reconciled with the [O\,II] interpretation.

\subsubsection{Interpretation as \OIII}
Should the line be \OIII~5007\AA, the redshift would be $z_{[OIII]}=2.20$ and $\rm EW_{OIII}\gtrsim2380$\AA. 
Large EW for \OIII\ have been observed \citep[e.g.,][]{boyett2022,tang2023,fujimoto2023,llerena2025,endsley2025}, potentially indicating the presence of a highly obscured active galactic nucleus (AGN), for which only the ``narrow line" region is detected (see section~\ref{sec:agn}).
However, if this line were \OIII\ $\lambda5007$, the accompanying \OIII\ $\lambda4959$ line should be detected at 15864~\AA. Based on the fixed doublet flux ratio, we would expect this line to have $S/N \approx 2.9$.
While this would not represent a strong detection, Figure~\ref{fig:medianline} shows no apparent feature at 15864~\AA, where we measure a $3\sigma$ upper flux limit of $\sim9.4\times10^{-19} ~ \rm erg \: s^{-1}cm^{-2}$. Additionally, no \Hb\ emission is detected at $1.55\,\mu\mathrm{m}$, where there is a $3\sigma$ line flux depth of $\sim9.2\times10^{-19} ~ \rm erg \: s^{-1}cm^{-2}$. Thus, the \OIII\ interpretation remains disfavored.

\subsubsection{Interpretation as \ha}
If this line were \ha, the galaxy would be at $z_{H\alpha}=1.44$ and the $\rm EW_{H\alpha}\gtrsim3130$\AA. This EW would be close to the theoretical maximum for a zero age starburst: for example, the maximal EW$^{\rm rest}_{\rm H\alpha}$ reached by a zero-age starburst is $\sim3200$\AA\ (5 Myr old population, with $Z=0.02Z_{\odot}$, and an IMF with $\alpha=1.5$ using Starburst99 models \citep{leitherer1999,shim2011}).
The measured line flux corresponds to a line luminosity of 
$3.3 \times 10^{40}\,\mathrm{erg\,s^{-1}}$. 
Using the \citet{kennicutt2012} calibration, this implies a star-formation rate of 
$\sim 0.2\,\mathrm{M_\odot\,yr^{-1}}$. Given the upper limit in the F115W flux, the stellar mass of this object must be below $\approx 10^6\mathrm{M_\odot}$, implying a specific SFR $\gtrsim 200\,\mathrm{Gyr^{-1}}$,
placing the source above the low-mass extrapolation of the star-forming main sequence at $z \sim 1.44$
\citep[e.g.,][]{noeske2007,nelson2016,merida2025}.
It's worth noting that we do not see \OIII\ 5007\AA\ at 12217\AA, where there is a $3\sigma$ line flux depth of $\sim9.5\times10^{-19} ~ \rm erg \: s^{-1}cm^{-2}$. There is also no \nii\ $6548,6583$\,\AA, \sii\ $6717,6731$\,\AA, \oi\ $6300$\AA\ and \hei\ $5876$\AA, all of which would appear in this wavelength range, but these lines would likely be much fainter than \ha.

The \OIII\ lines may or may not be intrinsically produced, as their strength 
depends sensitively on the ionizing spectrum and the gas-phase metallicity. 
If the starburst is effectively zero age (as implied by the large 
$\mathrm{EW}_{\mathrm{H}\alpha}$), the ionizing photon flux must be high, 
leading to a substantial O$^{+}$ photoionization rate. In this case, strong 
\OIII\ emission would generally be expected unless the metallicity is 
sufficiently low. A low metallicity could also account for the absence of 
detectable \nii\ and \sii\ emission, provided these lines are intrinsically 
weak and fall below the noise level.
However, a low nebular O/H abundance would imply minimal dust attenuation. 
In that scenario, \Hb\ should be observed close to its intrinsic Case~B value, 
yet no such emission is detected. This tension further disfavors an \ha\ 
interpretation for this emission line.

Contrary to metal lines, if the line is \ha, \Hb\ must be intrinsically produced. \Hb\ would be seen at 11865\AA\ is the line is \ha. Assuming Case~B recombination at $10^{4}\,\mathrm{K}$, the minimum intrinsic 
\Hb\ flux would be $8.6 \times 10^{-19}\,\mathrm{erg\,s^{-1}\,cm^{-2}}$, near the $3\sigma$ line flux upper limit of $\sim 9.7 \times 10^{-19}\,\mathrm{erg\,s^{-1}\,cm^{-2}}$ measured at this wavelength.
Assuming the same FWHM as \ha\ in velocity space, an \Hb\ line with this flux would be detected at a significance of $2.7\sigma$.
Although this would represent only a marginal detection, the spectrum shows no 
evidence of a feature at the expected wavelength.

There are two possibilities to weaken the \Hb\ flux with respect to the \ha. First, attenuation due to dust can significantly suppress the observed \Hb/\ha\ ratio, making a non-detection of \Hb\ plausible. The second option is that the Case~B approximation may not be an appropriate assumption. Some recent evidence suggests that  high values of the \ha/\Hb\ ratio could be due to large amount of dense excited gas, producing absorption in the Balmer transitions \citep{scarlata2024}.

\subsubsection{Interpretation as \lya}
Finally, another alternative is \lya, at a redshift of $z_{Ly\alpha}=12.2$ and a rest frame $\rm EW_{Ly\alpha}\gtrsim 580$\AA. If this line is \lya, then no other strong emission lines would fall within the G140M/F100LP spectral coverage, which would explain the lack of detection of other lines. \lya\ emission has been detected at even higher redshifts 
\citep[$z\sim13$, e.g., JADES-GS-z13-1-LA,][]{witstok2025}, demonstrating that such a scenario is not 
observationally implausible. 

However, JADES-GS-z13-1-LA differs substantially from the line-emitter from this work. JADES-GS-z13-1-LA galaxy exhibits a well-detected continuum redward of the 
Lyman break ($f_\nu \sim 6\,\mathrm{nJy}$ in F200W), corresponding to an 
absolute UV magnitude of $M_{\mathrm{UV}} \approx -18.5$. In contrast, our line-emitter lacks a comparable continuum detection. Furthermore, JADES-GS-z13-1-LA has a lower limit on the rest-frame equivalent width 
of $\gtrsim 40$~\AA, significantly smaller than the lower limit of 
$\gtrsim 580$~\AA\ inferred for this emission line. 
Such an extreme $\mathrm{EW}_{\mathrm{Ly}\alpha}$ would make this line-emitter
highly unusual. A rest-frame equivalent width of $\gtrsim 580$~\AA\ implies a 
very young and metal-poor stellar population \citep[e.g.,][]{schaerer2003,kashikawa2012}. 
For example, at zero age and $Z = 1/20\,Z_\odot$, \textsc{Starburst99} models 
can produce $\mathrm{EW}_{\mathrm{Ly}\alpha} \sim 500$~\AA\ for a Salpeter IMF 
with an extreme slope of $\alpha = 0.5$ \citep{malhotra2002}.

Additionally, interpreting the line as \lya\ would require minimal attenuation 
by the intergalactic medium (IGM), which would be remarkable at $z=12.16$.
At this redshift, the IGM is inferred to be $\gtrsim 90\%$ neutral 
\citep{bruton2023}, and ionized bubbles are not expected to be very extended 
\citep[e.g.,][]{hayesscarlata2023,umeda2023}, unless the sight-line happens to 
probe an unusually large ionized region. 
In the case of this line-emitter, its faintness implies that it cannot, by 
itself, ionize a sufficiently large bubble to allow \lya\ transmission. 
Sustaining such a bubble would instead require a substantial overdensity of 
nearby galaxies, such as a forming protocluster, capable of producing the 
necessary ionizing photon budget \citep{hu2021,lu2024}. Although none of the 
neighboring sources have photometric redshifts consistent with 
$z_{\mathrm{phot}} \sim 12.16$, protoclusters can extend over several physical 
Mpc and could, in principle, lie partially outside the current field of view.

Taken together, the extreme equivalent width requirement and the need for 
efficient \lya\ transmission through a highly neutral IGM both present 
significant challenges for a \lya\ interpretation of this line.

\subsection{Interpretation as an AGN}\label{sec:agn}
Given that star formation alone does not well explain this emission line, an active 
galactic nucleus (AGN) provides an alternative interpretation. In principle, 
any of the emission lines considered above could be powered by AGN activity. 
AGN typically reside in chemically enriched environments and are characterized 
by strong metal emission lines \citep[e.g.,][]{mulcahey2022}. If this line
were interpreted as \OIII\ or \ha, we would expect prominent accompanying 
metal lines within the observed wavelength range. In particular, in the \ha\ 
scenario, an enhanced \hei\ $\lambda5876$ recombination line would be expected 
\citep[e.g.,][]{almog1989}. The absence of such strong metal features therefore 
argues against a classical AGN interpretation.

In contrast, if the line were isolated \lya, the situation is somewhat  different. Rest-frame equivalent widths as large as 
$\sim 500$~\AA\ have been observed in faint quasars  \citep{matsuoka2019,harikane2022,lin2022}. 
However, the narrow width of this line places additional constraints on 
an AGN interpretation. The feature is unresolved in the MOCSS data at 
$R \sim 1000$, implying a velocity width of $\lesssim 300\,\mathrm{km\,s^{-1}}$. 
The emission-line could therefore arise from the narrow-line region of an AGN.

AGN variability can in principle produce artificially large inferred 
equivalent widths. For example, during a ``changing-look'' event, the hot 
accretion-disk continuum may fade rapidly while the line-emitting gas remains 
luminous for longer timescales \citep{komossa2024}, temporarily boosting the observed EW. 
Moreover, a rapid ``switch-on'' scenario is unlikely in this case: producing significant line 
variability on the $\sim 1$~yr timescale between the imaging and spectroscopic 
observations (see Section~\ref{sec:transient}) would require emission from the 
compact BLR, which is inconsistent with the observed narrow line width.

An alternative possibility is that an AGN located in one of the three 
neighboring sources listed in Table~\ref{tab:nearbysources} is powering the 
emission line. In this scenario, ionizing radiation from the AGN could 
illuminate a separate cloud of gas at the position of this line-emitter. 

This type of ``light echo'' configuration, analogous to systems such as 
``voorwerps'' and ``green beans'', allows for extremely large observed 
equivalent widths, since the ionizing source need not be co-spatial with the 
line-emitting gas \citep{lintott2009,trainorsteidel2013,cantalupo2014,davies2015,maseda2018,sanderson2024}. 
The neighboring sources to the northwest and south of the line-emitter's position have 
large uncertainties in their photometric redshifts, and their allowed 
$z_{\mathrm{phot}}$ ranges overlap with the redshifts at which this emission line 
could be interpreted as \ha, \OIII, or \OII. 
However, even in this externally illuminated scenario, the \OIII\ and \ha\ 
interpretations remain problematic, as they would still predict the presence 
of additional strong emission lines that are not observed.

\subsection{Interpretation as a Transient}\label{sec:transient}

In principle,  the emission line alone should be sufficiently bright to produce a 
detectable signal in the broad-band JADES F150W filter, even in the absence 
of any underlying continuum, so the lack of a corresponding imaging detection remains puzzling. 
To quantify this, we simulated the expected observed flux in F150W by generating 
synthetic photometry of the emission line using the \textsc{synphot} Python 
package \citep{synphot2018}. Convolving the emission-line spectrum with 
the NIRCam F150W transmission curve yields an expected flux density of 
$5.5\,\mathrm{nJy}$, corresponding to an observed magnitude of 
$m_{\mathrm{F150W}} \sim 29.6$, assuming no continuum contribution. 
This is approximately 0.7~mag brighter than the nominal JADES limiting 
magnitude, implying that the source should have been detected in the imaging 
data. The line flux of the emission line would need to be $\sim1.1\times10^{-18} \rm erg\ s^{-1} cm^{-2}$ or smaller to be absent in JADES imaging.
As discussed in Section~\ref{sec:spectralextraction}, the line flux may 
be overestimated by $\sim 3.3\%$, but such a correction is insufficient to 
reconcile the predicted flux with the non-detection.
This might indicate that a transient better explains the non detection in the imaging data. 

The MOCSS spectroscopic observations were obtained on 23 August 2023 
(orientation 1) and 19 November 2023 (orientation 2), while the JADES imaging 
used for comparison (Figure~\ref{fig:mysteryline}) was acquired on 
8–10 October 2022. Thus, there is approximately a one-year gap between the 
imaging and spectroscopic data.
In a transient scenario, the constraints on the observed equivalent width derived from the JADES non-detection would no longer apply, since the emission line may simply not have been present at the time of the imaging. 
Under this assumption, we take the maximum allowed continuum level to be $\sigma_{f_\lambda}$. In this case, the $3\sigma$ lower limit on the observed equivalent width is $\mathrm{EW}_{\rm obs} \gtrsim 41$~\AA. This corresponds to rest-frame limits of 
$\mathrm{EW}_{\mathrm{[O\,II]}} \gtrsim 10$~\AA, 
$\mathrm{EW}_{\mathrm{[O\,III]}} \gtrsim 13$~\AA, 
$\mathrm{EW}_{\mathrm{H}\alpha} \gtrsim 17$~\AA, and 
$\mathrm{EW}_{\mathrm{Ly}\alpha} \gtrsim 3$~\AA.
Therefore, in a transient interpretation, extremely large equivalent widths are no longer required to explain the emission line. However, transient scenarios introduce their own challenges in accounting for the observed properties of this emission line.

One possibility is a flaring star in the Milky Way (i.e., at $z=0$). 
Flaring stars can exhibit hydrogen Brackett-series transitions in emission \citep[e.g.,][]{howard2023}. In particular, the Brackett (13–4) transition at $1.61\,\mu\mathrm{m}$ lies close to the wavelength of the MOCSS emission line.  However, this interpretation predicts the presence of additional Brackett  lines, including Brackett-12 at $1.64\,\mu\mathrm{m}$, which is not observed. The absence of this feature effectively rules out a flaring-star scenario.

Another possibility is a Type~IIn supernova, which is characterized by narrow emission-line features \citep[e.g.,][]{taddia2013}. These narrow lines arise when the supernova blast wave shocks dense circumstellar material previously ejected by the progenitor through substantial mass loss.
The transient could be a supernova at $z \sim 0.25$, potentially associated with a low-surface-brightness disk component of one of the nearby JADES sources (two neighboring objects have best-fit photometric redshifts near $z \sim 0.25$). In this scenario, this emission line could be identified as \pab. However, this interpretation predicts the presence of \pag\ at $1.37\,\mu\mathrm{m}$. Assuming Case~B recombination, the expected flux ratio is $\mathrm{Pa}\beta/\mathrm{Pa}\gamma \sim 1.6$, implying that \pag\ should be detected at $>3\sigma$ significance. That said, supernovae make a lot of dust, which could preferentially attenuate the \pag\ line and make it appear weaker.

Another possible transient explanation is a tidal disruption event (TDE). A TDE occurs when a supermassive black hole tidally disrupts a star, producing a luminous flare \citep{kool2020,cannizzaro2022,leloudas2022}. At late times, 
TDE spectra can exhibit strong \ha\ emission accompanied by weak continuum, as well as a wide range of observed \ha/\Hb\ line ratios 
\citep{charalampopoulos2022}. In luminous, dust-obscured TDEs, optically thick gas can suppress forbidden emission lines such as \OIII, while dust attenuation can reduce the strength of \Hb\ \citep{panagiotou2023}. In such a scenario, \ha\ could plausibly be observed in isolation within the wavelength range, without detectable \OIII\ or \Hb. 

The two MOCSS orientations were observed at separate epochs approximately 88 days apart, yet the measured line fluxes are consistent between the two observations. While this does not exclude a transient origin—since some TDEs evolve on longer timescales—it does argue against a rapidly moving solar system object.

\section{Conclusion}\label{sec:conclusion}
In this study, we report the identification of a unique emission line using a new NIRSpec observing strategy, Multi-Object Coronagraphy for Slitless Spectroscopy (MOCSS). This technique involves opening nearly all shutters in the NIRSpec MSA while selectively closing those that coincide with known galaxies. This approach enabled high-resolution ($R \sim 1000$) spectroscopy over $0.97$–$1.89\,\mu\mathrm{m}$ in the HUDF, 
resulting in the discovery of a strong emission line at $1.6\,\mu\mathrm{m}$.

The line is robustly detected in multiple dithers and in two independent orientations, with a signal-to-noise ratio of 5.4 and no detectable continuum. 
There is no corresponding source in existing JADES NIRCam imaging. Based on this non-detection, the inferred observed equivalent width is extremely large ($\mathrm{EW}_{\rm obs} \sim 7620$~\AA), implying that, if real and persistent, 
the source would have to be a very young and metal-poor galaxy.

However, none of the plausible line identifications, \OII, \OIII, \ha, or \lya, provide a fully satisfactory explanation, whether or not an AGN is invoked. 
Transient scenarios also face significant challenges, although they would naturally explain the lack of an imaging counterpart and would relax the extreme equivalent width constraints.

Regardless of its true nature—whether an extremely low-metallicity, primeval galaxy at $z \sim 12.2$, a distant AGN in a fading phase, or an extreme metal-poor source at $z \sim 1.4$—this object is highly unusual and probes extreme physical conditions.

Follow-up observations are essential to confirm the redshift and determine the physical origin of the emission. JWST/NIRSpec integral field unit (IFU) observations would be particularly powerful, enabling simultaneous spectroscopy of both this line-emitter and nearby galaxies. Deeper, more sensitive  spectroscopy could reveal additional emission lines and definitively establish 
the redshift and nature of the source.

\begin{acknowledgments}
We thank the JWST help desk for providing substantial feedback and assistance with the data reduction. MSH thanks Annalisa Citro and Alexandra Le Reste for providing useful tools to assist with this analysis. 
We acknowledge support by NASA
through grant JWST-GO-3290. All the JWST data used in this paper can be found in MAST: \dataset[10.17909/tct8-jr92]{http://dx.doi.org/10.17909/tct8-jr92}.
MJH is supported by the Swedish Research Council (Vetenskapr{\aa}det) and is Fellow of the Knut \& Alice Wallenberg Foundation. VC thanks the BlackHoleWeather project and PI Prof. Gaspari for salary support. MSH thanks Prof. Gaspari for useful discussion.

\software{NumPy \citep{Numpy:2020}; SciPy \citep{Scipy:2020}; AstroPy \citep{Astropy:2013,Astropy:2018,Astropy:2022}; Matplotlib \citep{Matplotlib:2007}.}

\end{acknowledgments}




\bibliography{sample701}{}
\bibliographystyle{aasjournalv7}



\end{document}